\documentclass[11pt,twoside]{article}
\usepackage{asp2010}

\newcommand{\hi}{{\rm H{\textsc{i}}\,}}

\resetcounters

\begin{document}

\title{E(A+M)PEC - An OpenCL Atomic \& Molecular Plasma Emission Code For Interstellar Medium
Simulations}
\author{Miguel A. de Avillez$^{1}$, Emanuele Spitoni$^{1}$, \& Dieter Breitschwerdt$^{2}$
\affil{$^1$Dept. Mathematics, U. \'Evora, R. Rom\~ao Ramalho 59, 7000 \'Evora, Portugal}
\affil{$^2$ZAA, Technische Universit\"at Berlin, Hardenbergstr.~36, Berlin, Germany}
}

\begin{abstract}
E(A+M)PEC traces the ionization structure, cooling and emission spectra of plasmas. It is written in
OpenCL, runs in NVIDIA Graphics Processor Units and can be coupled to any HD or MHD code to follow the
dynamical and thermal evolution of any plasma in, e.g., the interstellar medium (ISM).
\end{abstract}
\section{Introduction}
The thermal evolution of the ISM is determined by heating and cooling processes which may not be synchronized with
ionization and recombination processes and the system deviates from collisional ionization equilibrium (CIE). A step
forward in ISM modelling requires that both the thermal and dynamical evolutions be treated in a self-consistent way and
under non-equilibrium ionization conditions (NEI).
\begin{figure}[thbp]
\centering
\plottwo{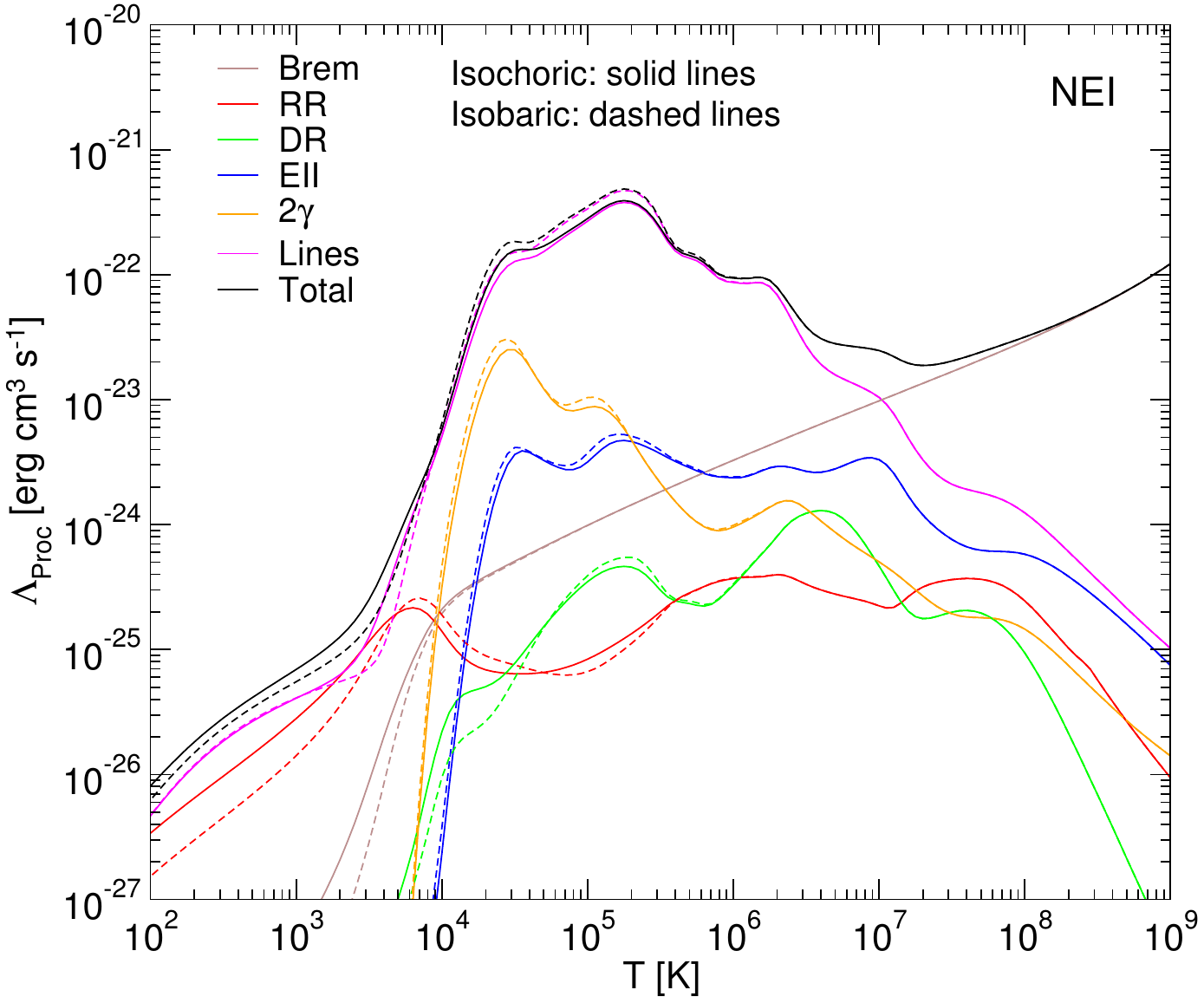}{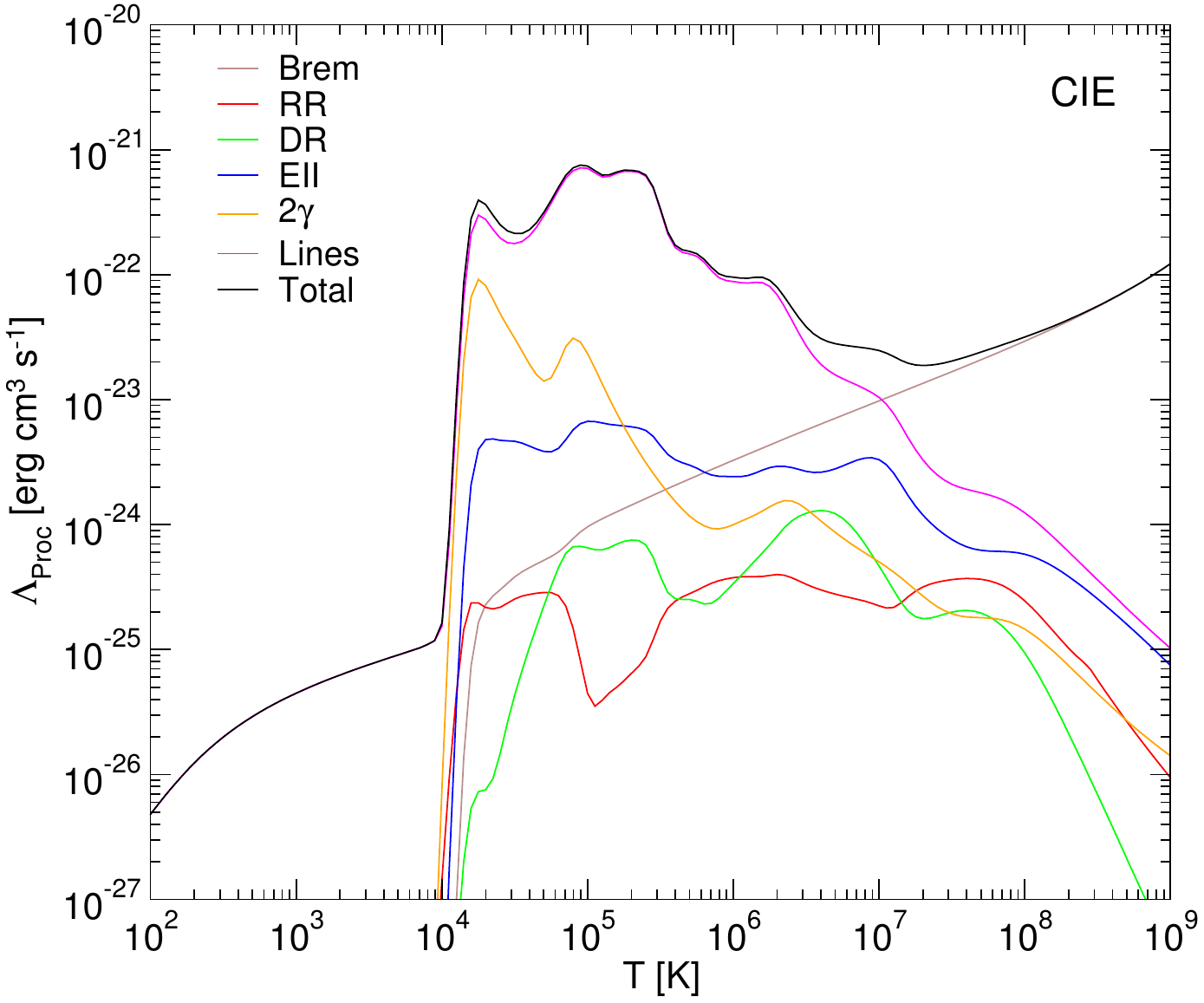}
\caption{\small Normalized to n$_{H}^{2}$ NEI (left: solid lines and dashed lines show cooling under isochoric and
isobaric coditions, respectively) and CIE (right) cooling efficiencies using \citet{agss2009}
solar abundances. Total cooling (black lines), free-free (brown line), radiative (red) and dielectronic
(green) recombination, electron impact ionization (blue), two-photon ($2\gamma$; orange) and line excitation
excluding $2\gamma$ (magenta) emissions.}
\label{cooling}
\end{figure}

\section{E(A+M)PEC Features and Some Results}

\begin{figure}[thbp]
\centerline{\includegraphics[width=0.6\hsize,angle=0]{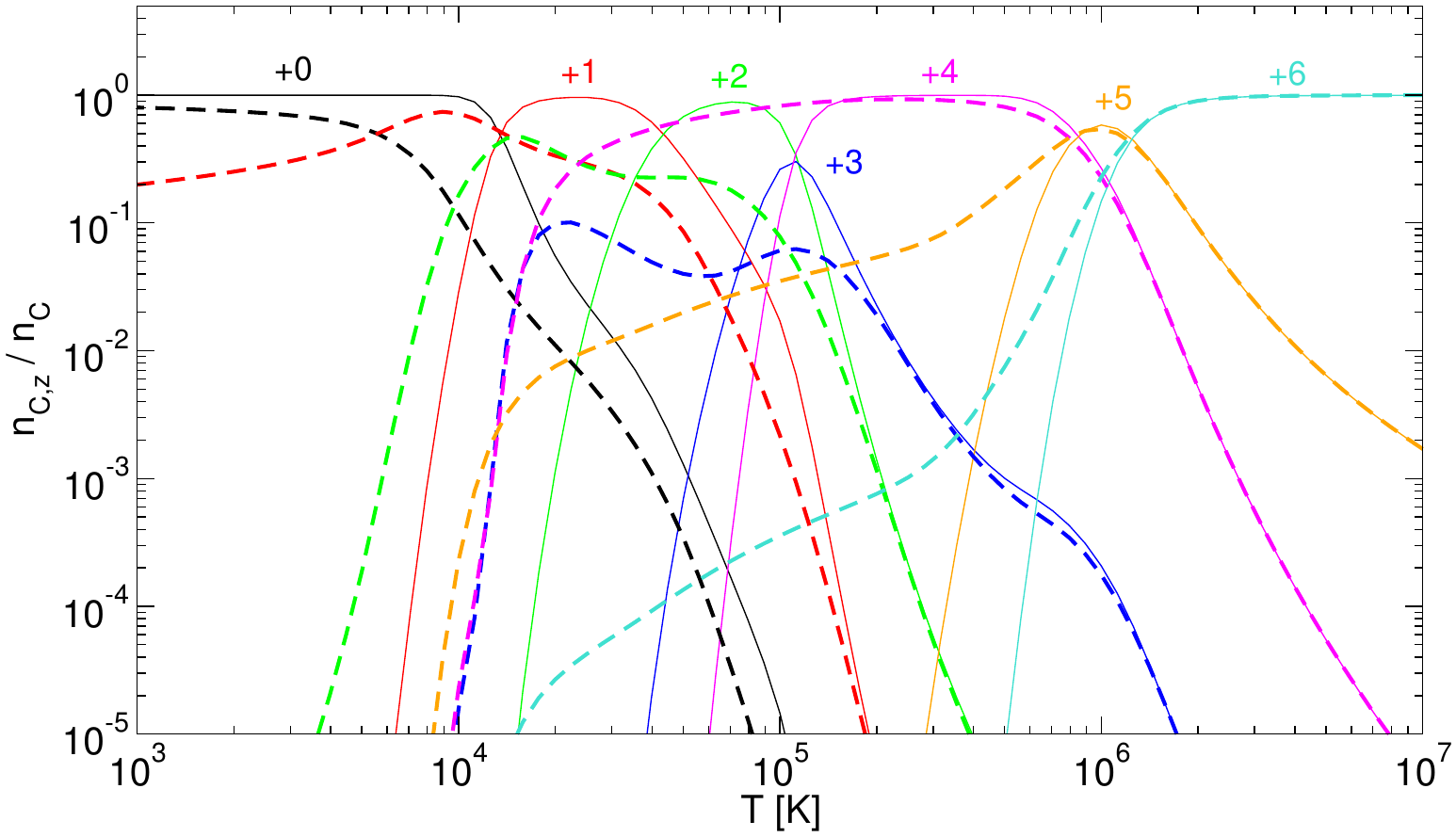}}
\caption{\small CIE (solid lines) and NEI isochoric (dashed lines) ionization structure of Carbon between
$10^{2}$ and $10^{7}$ K. Charge-exchange reactions have been included.}
\label{carbon}
\end{figure}

E(A+M)PEC traces the ten most abundant elements in nature, includes up to date solar abundances and atomic and
molecular data (see e.g., http://www.lca.uevora.pt/research and references therein). The physical processes
comprise electron impact ionization, excitation auto-ionization, radiative and dielectronic recombination,
charge-exchange reactions, ionization of \hi as result of He ions recombination, continuum (free-free,
free-bound, two-photon) and line mission ($\lambda \in [1\AA,~610\mu]$), and H, C and O chemistry,

The cooling efficiencies of a plasma evolving from $10^{9}$ K (where it was fully ionized) to 100 K under CIE
and NEI (isochoric and isobaric) conditions are shown in Figure~\ref{cooling}. Departure from CIE occurs at
some $10^{6}$ K. As recombination lags behind, single and double ionized species exist at lower temperatures
(Figure~\ref{carbon}) something that does not occur under CIE (because ionization and recombination is
synchronized) and therefore neutrals form at temperatures near $10^{4}$ K.
These differences also affect the emission spectra - with recombination lagging behind, typically around
$10^{5}$ K, the spectra at $\lambda < 100$\AA~ becomes free-bound dominated under NEI conditions.

\section{Final Remarks}

The ionization structure, cooling and emission spectra obtained with the E(A+M)PEC running in NVIDIA GPUs
are similar (though with some deviations due to different atomic data and abundances) to results of, e.g,
\citet{st1993}, \citet{gs2007} and \citet{bls2009}. Regarding ISM simulations, it should be stressed that the
interstellar cooling function is changing in space and time (Avillez \& Breitschwerdt, submitted).

\acknowledgements 

Research funded by FCT project PTDC/CTE-AST/70877/2006.

\bibliography{author}

\end{document}